\documentclass[aps,pra,twocolumn,showpacs,amsmath,amssymb]{revtex4-1}

\usepackage[utf8]{inputenc}

\usepackage{graphicx}
\usepackage[colorlinks=true,citecolor=blue]{hyperref}
\usepackage{units}

\usepackage{verbatim} %mehrzeilige Kommentare
\usepackage{abbrevs}

\newcommand{\kBT}{k_\text{B}T}

\def\kB {k_\text{B}}

\begin{document}

\title{Quantum heat engines based on electronic Mach-Zehnder interferometers}

\author{Patrick P. Hofer}
\affiliation{Département de Physique Théorique, Université de Genève, CH-1211 Genève 4, Switzerland}
\author{Björn Sothmann}
\affiliation{Département de Physique Théorique, Université de Genève, CH-1211 Genève 4, Switzerland}

\date{\today}

\begin{abstract}
We theoretically investigate the thermoelectric properties of heat engines based on Mach-Zehnder interferometers. The energy dependence of the transmission amplitudes in such setups arises from a difference in the interferometer arm lengths. Any thermoelectric response is thus of purely quantum mechanical origin. 
% In addition to an experimentally established three-terminal setup, we investigate two improvements. A two-terminal setup, corresponding to different boundary conditions employed on the three-terminal setup, and a four-terminal setup consisting of two interferometers.
In addition to an experimentally established three-terminal setup, we also consider a two-terminal geometry as well as a four-terminal setup consisting of two interferometers.
We find that Mach-Zehnder interferometers can be used as powerful and efficient heat engines which perform well under realistic conditions.
\end{abstract}

% \pacs{73.23.Hk,72.70.+m,72.25.Mk,85.75.-d}
% 73.23.Hk 	Coulomb blockade; single-electron tunneling
% 73.23.-b 	Electronic transport in mesoscopic systems
% 73.63.-b 	Electronic transport in nanoscale materials and structures
% 72.70.+m 	Noise processes and phenomena
% 73.63.Kv 	Quantum dots
% 72.25.Mk 	Spin transport through interfaces
% 85.75.-d 	Magnetoelectronics; spintronics: devices exploiting spin polarized transport or integrated magnetic fields

%\keywords{}

\maketitle

\section{\label{sec:Intro}Introduction}
Thermoelectric effects have received a lot of attention in recent years because they open up the possibility to recover waste heat back into useful electricity~\cite{shakouri_recent_2011}. Mesoscopic conductors are potentially promising candidates to achieve this goal. Nanostructured materials such as quantum wires and wells~\cite{hicks_thermoelectric_1993,hicks_effect_1993} have been shown to exhibit an increased thermoelectric figure of merit. The increased thermoelectric performance of nanoscale conductors can be traced back to their sharp spectral features that allows for efficient energy filtering~\cite{mahan_best_1996}.
So far, most studies, both on the experimental~\cite{molenkamp_quantum_1990,staring_coulomb-blockade_1993} as well as on the theoretical~\cite{streda_quantised_1989,beenakker_theory_1992} side were limited to two-terminal devices. 
More recently, multi-terminal thermoelectrics generated a lot of interest~\cite{entin-wohlman_three-terminal_2010,sanchez_optimal_2011,sothmann_rectification_2012,entin-wohlman_three-terminal_2012,ruokola_theory_2012,sothmann_magnon-driven_2012,jordan_powerful_2013,sothmann_powerful_2013,jiang_three-terminal_2013,machon_nonlocal_2013,bergenfeldt_hybrid_2014,mazza_thermoelectric_2014,sothmann_quantum_2014,sanchez_chiral_2014}, for a recent review see Ref.~\cite{sothmann_thermoelectric_2015}. 
These setups offer the advantage of crossed flows of heat and charge currents. This feature allows for an electrical separation of the hot source from the actual rectifier which is a crucial prerequisite for energy harvesting applications.

During the past few years, there has been an interest in thermoelectric effects in the absence of time-reversal symmetry. In this situation, the off-diagonal Onsager coefficients that fully characterize thermoelectric transport in the linear response regime do not have to be equal to each other. This additional degree of freedom opens up the possibility to have (at least in principle) Carnot efficiency $\eta_C$ at finite output power~\cite{benenti_thermodynamic_2011}. However, in multi-terminal setups, current conservation imposes additional constraints on the scattering matrix that reduces the efficiency at maximum power $\eta_\text{maxP}$ to a value smaller than Carnot efficiency~\cite{brandner_strong_2013,brandner_multi-terminal_2013}. At the same time the bound $\eta_\text{maxP}\leq\eta_C/2$ derived by van den Broeck~\cite{van_den_broeck_thermodynamic_2005} for time-reversal symmetric systems can be broken. So far, there have been theoretical studies of thermoelectrics with broken time-reversal symmetry in triple quantum dots~\cite{saito_thermopower_2011}, classical Nernst engines~\cite{stark_classical_2014}, quantum Nernst engines~\cite{sothmann_quantum_2014} and in chiral heat engines based on quantum Hall edge states~\cite{sanchez_chiral_2014}.

\begin{figure}
	\includegraphics[width=\columnwidth]{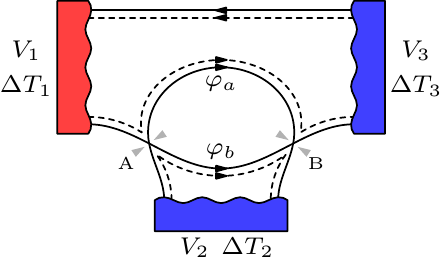}
	\caption{\label{fig:setup1}Schematic sketch of a three-terminal heat engine based on an electronic Mach-Zehnder interferometer. By heating contact $1$, a charge current is carried by quantum Hall edge states (solid lines) between contacts $2$ and $3$ due to the energy dependence of the transmission amplitudes. In the MZI, it is the phase-difference $\varphi_a-\varphi_b$ which provides the energy dependence due to a difference in the path lengths. The dashed lines represent a second edge channel at filling factor two. Note that in an experimental setup, contact $2$ has to be fabricated within the interferometer. Here, $A$ and $B$ denote quantum-point contacts, $V_\alpha$ and $\Delta T_\alpha$ the bias voltage and the temperature bias applied at contact $\alpha$.}
\end{figure}

Here, we investigate a quantum heat engine based on an electronic Mach-Zehnder interferometer~\cite{ji_electronic_2003} (MZI) in a two-dimensional electron gas in the quantum Hall regime (see Fig.~\ref{fig:setup1}). Within this interferometer, the electronic wave-functions are delocalized over two paths which enclose a magnetic flux. This flux modifies the difference of the phases picked up along the two paths. The resulting flux-dependent oscillations in the current were measured with visibilities reaching over $80\%$~\cite{neder_interference_2007}. Due to this high degree of coherence, the MZI has been employed successfully in a number of experiments investigating coherent electron transport~\cite{ji_electronic_2003,neder_interference_2007,neder_controlled_2007,neder_entanglement_2007,roulleau_direct_2008,roulleau_noise_2008,roulleau_tuning_2009,litvin_phase_2010,helzel_counting_2012,huynh_quantum_2012}.

The main motivation for our work derives from the fact that in a MZI the energy dependence of the scattering matrix arises from the path length difference between the two interferometer arms only. Thus, any thermoelectric response is of purely quantum mechanical origin. Furthermore, as time-reversal symmetry is broken by the strong magnetic field required for the quantum Hall effect, it offers an interesting playground to study the effects of broken time-reversal symmetry on the thermoelectric response in a mesoscopic multi-terminal heat engine.

The rest of the paper is structured as follows. In Sec.~\ref{sec:MZI}, we introduce the MZI as well as the important thermoelectric quantities that are used throughout the text. Our results for three different setups are presented in Sec.~\ref{sec:results}. Before concluding in Sec.~\ref{sec:Conclusions}, we discuss the experimental feasibility of our proposal in Sec.~\ref{sec:exp}.

\section{\label{sec:MZI}Heat engines based on Mach-Zehnder interferometers}
We describe our system with a non-interacting scattering matrix approach, which is suitable to describe the linear response regime of interest even for a MZI, where interaction effects are known to be of relevance in general~\cite{neder_controlled_2007}. The transmission probabilities through a MZI are then of the form
\begin{equation}
\label{eq:transmzi}
\left|A_ae^{i\varphi_a}+A_be^{i\varphi_b}\right|^2=A_a^2+A_b^2+2A_aA_b\cos\varphi,
\end{equation}
where $A_i$ is the probability amplitude for going along path $i$ and $\varphi_i$ is the phase picked up along that path. Linearizing the spectrum $E=\hbar v_D(k-k_F)$, the energy-dependent difference of these phases is given by
\begin{equation}
\label{eq:phasemzi}
\varphi=\varphi_a-\varphi_b=\phi+E\tau/\hbar.
\end{equation}
Here $\phi$ is the magnetic flux in units of $\hbar/e$ enclosed by the two paths and $\tau=(L_a-L_b)/v_D$ is the length-difference of the paths divided by the drift velocity. Constant phases arising from the QPC's and the chemical potential are absorbed in $\phi$. The transmission probabilities through a MZI thus show oscillations as a function of energy for an imbalanced MZI ($L_a\neq L_b$). It is this energy dependence which allows the MZI to be operated as a heat engine. The effects of a path-length difference on the charge current and noise has been investigated in Ref.~\onlinecite{chung_visibility_2005} for a dc bias and in Ref.~\onlinecite{hofer_mach-zehnder_2014} for an ac bias. Since we have
\begin{equation}
\label{eq:a1a2}
A_aA_b\propto\sqrt{\mathcal{R}_A\mathcal{D}_A\mathcal{R}_B\mathcal{D}_B},
\end{equation}
where $\mathcal{R}_i$ and $\mathcal{D}_i$ are reflection and transmission probabilities of the quantum-point contact (QPC) $i$, we focus on MZIs with half-transparent QPCs (i.e. $\mathcal{R}_i=\mathcal{D}_i=1/2$), where the interference effects are most pronounced. Note that a finite temperature, in combination with an energy-dependent transmission probability \textit{decreases} the interference effect due to phase averaging. This determines a finite window for the product of temperature and path length difference where a thermally induced charge current is created exploiting the interference effect. Additional sources of dephasing will be discussed below.

We now introduce the different quantities that characterize the performance of a given setup as a heat engine. To constitute a heat engine, the system must produce a charge current $I^e$ as a response to a thermal bias $\Delta T$ between the hot and the cold reservoirs. In order to perform work, this thermal current has to overcome an externally applied voltage bias $V$. 
The efficiency with which the heat engine converts thermal energy into electrical energy is given by the ratio between the output power and the input power. The input power is equal to the heat current $I^h$ which enters the system through the hot contacts and the output power reads
\begin{equation}
\label{eq:power}
P=-I^eV.
\end{equation}

Throughout the paper, the quantities without a contact index refer to the currents and forces that characterize a given setup as a heat engine. Having identified these quantities, it is convenient to relate them via the reduced Onsager matrix~\cite{brandner_strong_2013,brandner_multi-terminal_2013}
\begin{equation}
	\left(\begin{array}{c}I^e\\I^h\end{array}\right)
	=
	\left(\begin{array}{cc} \mathcal L^{eV} & \mathcal L^{eT} \\ \mathcal L^{hV} & \mathcal L^{hT}\end{array} \right)
	\left(\begin{array}{c}F^V\\F^T\end{array}\right),
\end{equation}
where the thermodynamic forces are given by
\begin{equation}
\label{eq:tdforces}
F^V=\frac{eV}{k_B T},\hspace{.5cm}F^T=\frac{k_B\Delta T}{(k_B T)^2}.
\end{equation}
The coefficients of the reduced Onsager matrix then encode all the information on the heat engine in the linear response regime. For instance, the Seebeck coefficient, which gives the ratio between bias voltage and temperature bias at which the current vanishes reads
\begin{equation}
\label{eq:seebeck}
S=-\left.\frac{V}{\Delta T}\right|_{I^e=0}=\frac{\mathcal{L}^{eT}}{eT\mathcal{L}^{eV}}.
\end{equation}
Similarly, the power maximized with respect to the applied bias voltage is given by
\begin{equation}
\label{eq:maxp}
P_{\rm max}=\frac{k_BT}{4e}\frac{\left(\mathcal{L}^{eT}\right)^2}{\mathcal{L}^{eV}}\left(F^T\right)^2,
\end{equation}
and the efficiency at maximum power by
\begin{equation}
\label{eq:effmax}
\begin{aligned}
\eta_{\rm maxP}=\frac{P_{\rm max}}{I^h}=&\frac{\eta_C}{2e}\frac{\left(\mathcal{L}^{eT}\right)^2}{2\mathcal{L}^{eV}\mathcal{L}^{hT}-\mathcal{L}^{eT}\mathcal{L}^{hV}}\\
=&\eta_C\frac{\sqrt{1+ZT}-1}{\sqrt{1+ZT}+1},
\end{aligned}
\end{equation}
where $\eta_C=\Delta T/T$ denotes the Carnot efficiency in linear response. In the second line of the last expression, we give the relation between $\eta_{\rm maxP}$ and the figure of merit $ZT$. Since these quantities contain the same information, in the following we will describe the proposed heat engines using $\eta_{\rm maxP}$.

\section{\label{sec:results}Results}

In its most simple implementation, the MZI is usually fabricated as a three-terminal setup (see, e.g., Ref.~\onlinecite{ji_electronic_2003}) consisting of a source and two drains. Figure \ref{fig:setup1} schematically depicts such a three-terminal setup with contact $1$ being the source. We will first discuss this experimentally established setup and show that it can be employed as a heat engine with moderate efficiency in Sec.~\ref{ssec:3terminal}. In Sec.~\ref{ssec:2terminal}, we show how the efficiency can be improved by employing different boundary conditions which results in an effective two-terminal setup. Finally, in Sec.~\ref{ssec:4terminal}, we analyze a four-terminal heat engine based on two Mach-Zehnder interferometers which constitutes our most powerful and efficient setup.

\subsection{\label{ssec:3terminal}Three-terminal MZI}
For the three-terminal setup sketched in Fig.~\ref{fig:setup1}, the transmission probabilities read (for half-transparent QPCs)
\begin{equation}
\label{eq:3smat}
\begin{aligned}
&\left|S_{13}\right|^2=1,\\& \left|S_{11}\right|^2=\left|S_{12}\right|^2=\left|S_{33}\right|^2=\left|S_{23}\right|^2=0,\\
&\left|S_{31}\right|^2=\left|S_{22}\right|^2=\frac{1}{2}\left(1-\cos\varphi\right),\\
&\left|S_{32}\right|^2=\left|S_{21}\right|^2=\frac{1}{2}\left(1+\cos\varphi\right),
\end{aligned}
\end{equation}
where $\varphi$ is given by Eq.~\eqref{eq:phasemzi} and $S_{\alpha\beta}$ is the transmission amplitude from contact $\beta$ to contact $\alpha$. In order to operate the system as a heat engine, we heat up contact $1$ and we demand a vanishing charge current in the hot contact $I^e_1=0$. The current produced by the engine then flows between the cold contacts. Since the only channel entering contact $1$ is coming from contact $3$ with unit transmission, we find $V_1=V_3$. Together, these considerations yield
\begin{equation}
\label{eq:3tvs}
\begin{aligned}
&T_1=T+\Delta T, \hspace{0.25cm}T_2=T_3=T,\hspace{0.25cm}V_1=V_3=0,\\
&I^e=I^e_3,\hspace{0.25cm}I^h=-I^h_1,\hspace{0.25cm}V_2=V,
\end{aligned}
\end{equation}
The charge and heat currents can be calculated using the standard expressions (see e.g., Ref.~\cite{moskalets:book})
\begin{equation}
\label{eq:currentcharge}
I^e_\alpha=\frac{e}{h}\sum_{\beta}\int\limits_{-\infty}^{\infty}dE\left|S_{\alpha\beta}\right|^2[f_\beta(E)-f_\alpha(E)],
\end{equation}
and
\begin{equation}
\label{eq:currentheat}
I^h_\alpha=\frac{1}{h}\sum_{\beta}\int\limits_{-\infty}^{\infty}dE(E-eV_\alpha)\left|S_{\alpha\beta}\right|^2[f_\beta(E)-f_\alpha(E)],
\end{equation}
expanded to linear order in bias voltage and temperature bias.
The Fermi distribution in contact $\alpha$ is given by
\begin{equation}
\label{eq:fermi}
f_\alpha(E)=\frac{1}{e^{(E-eV_\alpha)/k_BT_\alpha}+1}.
\end{equation}
For the Onsager coefficients we then find
\begin{align}
\label{eq:3onsev}
	\mathcal L^{eV}&=\frac{ek_BT}{2h}\left[1+s(z)\cos\phi\right],\\
	\label{eq:3onset}
	\mathcal L^{eT}&=\frac{e\pi (k_BT)^2}{2h}g(z)\sin\phi,\\
	\label{eq:3onshv}
	\mathcal L^{hV}&=0,\\
	\label{eq:3onsht}
	\mathcal L^{hT}&=\frac{\pi^2(\kBT)^3}{3h},
\end{align}
with the auxiliary functions 
\begin{equation}
\label{eq:auxfcs}
s(z)=\frac{z}{\sinh z},\hspace{.5cm}g(z)=\frac{z\coth z-1}{\sinh z},
\end{equation}
and the abbreviation
\begin{equation}
\label{eq:z}
z=\frac{\pi \kBT \tau}{\hbar}.
\end{equation}

From Eq.~\eqref{eq:3onset}, we see that the current induced by the temperature bias indeed only has an interference term which oscillates with $\phi$. As can be seen from Eq.~\eqref{eq:3onsev}, the current driven by the voltage bias has both a constant as well as an interference part. Furthermore, while we find a finite Seebeck effect (i.e. thermally induced charge current or finite $\mathcal L^{eT}$) the Peltier effect vanishes (i.e. no electrically induced heat current or vanishing $\mathcal L^{hV}$). Since time-reversal symmetry enforces $\mathcal L^{eT}=e\mathcal L^{hV}$~\cite{onsager_reciprocal_1931,casimir_onsagers_1945}, this extreme asymmetry of the reduced Onsager matrix is a direct consequence of the broken time-reversal symmetry in quantum Hall systems. In this case, the maximal efficiency equals the efficiency at maximum power (because the heat current is independent of the bias voltage) and is bounded from above by $\eta_{\rm maxP}\leq\eta_C/4$ for a three-terminal setup~\cite{brandner_strong_2013}.

Evaluating Eq.~\eqref{eq:seebeck}, we find for the Seebeck coefficient
\begin{equation}
\label{eq:3seebeck}
S=\frac{\pi k_B}{e}\frac{g(z)\sin\phi}{1+s(z)\cos\phi},
\end{equation}
which reaches up to $S\approx1.26\frac{\kB}{e}\approx\unit[108]{\mu V/K}$ at maximal power (see below). A slightly higher value of $S\approx1.81\frac{\kB}{e}\approx\unit[156]{\mu V/K}$ can be reached for $z\rightarrow0$ and $\phi=\pi$. However, the delivered power vanishes in this configuration. These values are comparable to conventional bulk semiconductor thermoelectric devices~\cite{shakouri_recent_2011}. We note that at $\phi=0$, the thermoelectric response vanishes. This is a consequence of the linear dispersion for which the transmission probabilities become symmetric around the Fermi energy in the absence of an Aharonov-Bohm flux restoring particle-hole symmetry[cf.~Eqs.~\eqref{eq:phasemzi} and~\eqref{eq:3smat}].

\begin{figure}
	\includegraphics[width=\columnwidth]{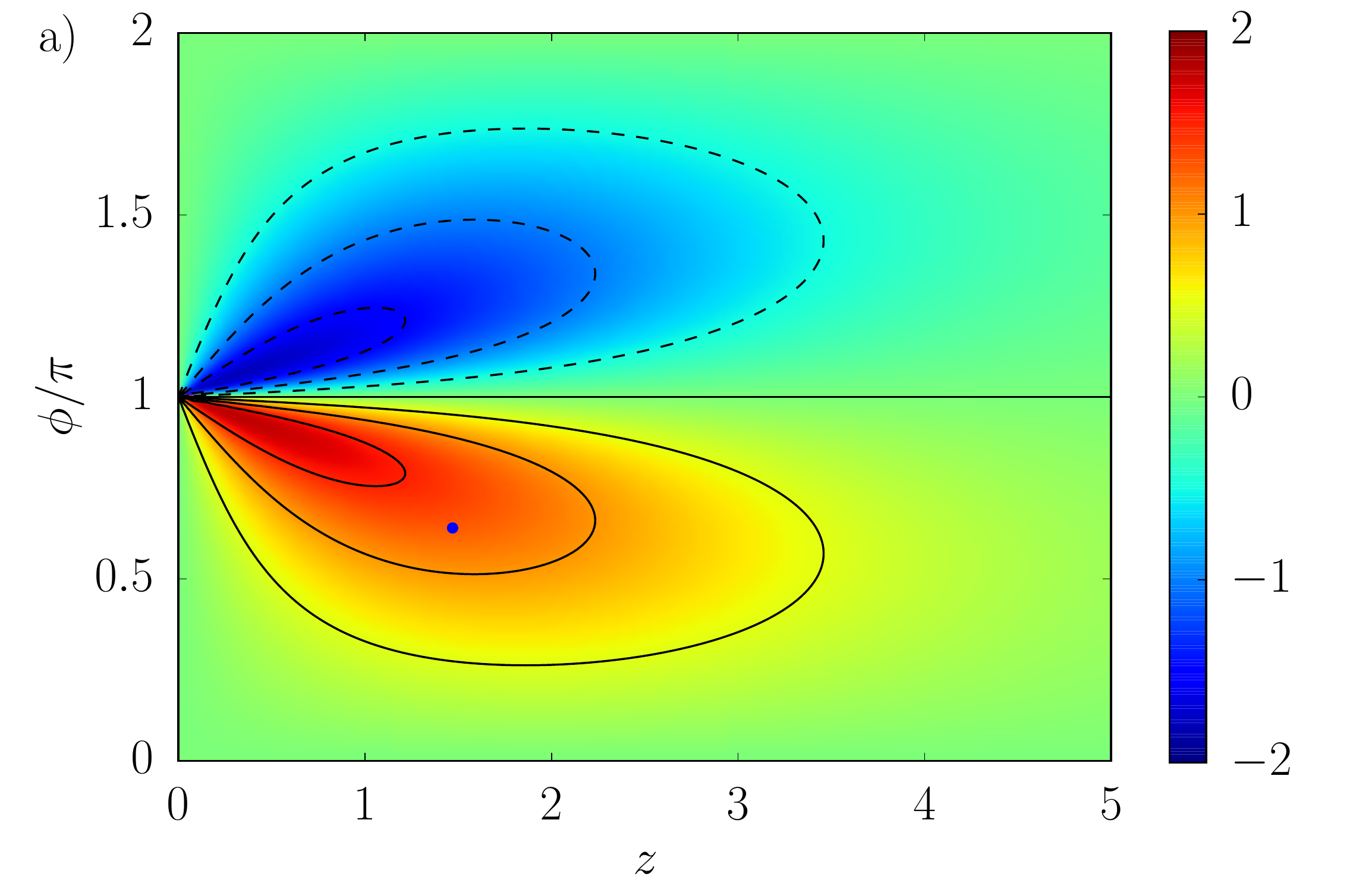}
	\includegraphics[width=\columnwidth]{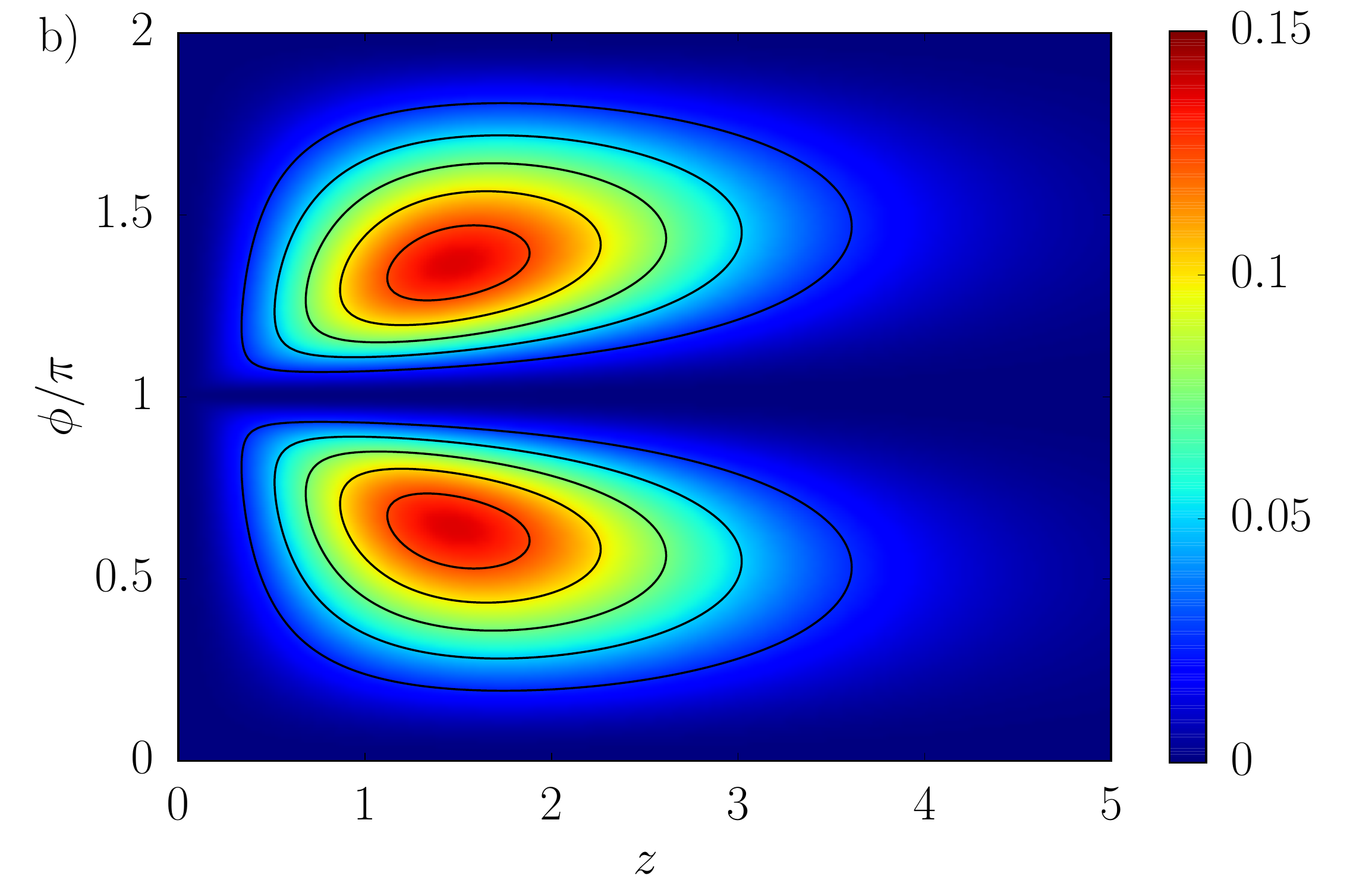}
	\caption{\label{fig:seebeckpower3}Seebeck coefficient and maximum power for the three-terminal setup. (a) Seebeck coefficient in units of $\kB/e$, the blue dot denotes the point where the power is maximized. (b) Power in units of $(\kB\Delta T)^2/h$.}
\end{figure}

The maximal power evaluates to %[cf.~Eq.~\eqref{eq:maxp}]
	\begin{equation}
	\label{eq:3maxp}
		P_\text{max}=\frac{(\pi\kB\Delta T)^2}{8h}\frac{g^2(z)\sin^2\phi}{1+s(z)\cos\phi},
	\end{equation}
which is maximized for $z\approx 1.47$ and $\phi\approx 0.64\pi$, reaching approximately $0.14(k_B\Delta T)^2/h=\unit[0.04]{pW/K^2}(\Delta T)^2$. This is an increase in output power by one order of magnitude compared to heat engines based on chaotic cavities~\cite{sothmann_rectification_2012} and only a factor of three less than for heat engines based on resonant-tunneling quantum dots~\cite{jordan_powerful_2013}.
The Seebeck coefficient as well as the delivered power are plotted in Fig.~\ref{fig:seebeckpower3}.

The efficiency at maximum power is given by %[cf.~Eq.~\eqref{eq:effmax}]
	\begin{equation}
	\label{eq:3effmax}
		\eta_\text{maxP}=\eta_C\frac{3g^2(z)\sin^2\phi}{8[1+s(z)\cos\phi]},
	\end{equation}
where the Carnot efficiency in linear response reads $\eta_C=\Delta T/T$. Since the heat current is independent of the MZI [due to the direct connection from contact $3$ to $1$, see also Eq.~\eqref{eq:3onsht}], the efficiency is maximized together with the power and reaches up to $\eta_\text{maxP}=0.042\eta_C$ which is much smaller than the theoretical bound of $\eta_C/4$~\cite{brandner_strong_2013}.

In the following subsections, we discuss two ways of increasing the efficiency of the proposed heat engine. For any setup based on a MZI, we find that the maximal power will be of a form similar to Eq.~\eqref{eq:3maxp} [see Eqs. (\ref{eq:doublepower},~\ref{eq:doublepowernu2}) below]. This is a direct consequence of the fact that the sole energy dependence of the MZI enters through the phase-difference $\varphi$. However, in the three-terminal setup discussed above, the heat current in the hot contact has no interference contribution. This is a consequence of the unit transmission from contact $3$ to contact $1$. To increase the efficiency of the heat engine, we can thus try to reduce the heat current $I^h$. Below we discuss how this can be achieved by either combining contacts $1$ and $3$ (obtaining a two-terminal setup, see Sec.~\ref{ssec:2terminal}) or by inserting an additional MZI in between contacts $1$ and $3$ (obtaining a four terminal setup, see Sec.~\ref{ssec:4terminal}). Before embarking on this investigation, we conclude this section with a brief discussion on the influence of dephasing and higher filling factors.

To introduce dephasing, we couple one of the arms of the interferometer to a fictitious voltage probe~\cite{chung_visibility_2005}. The effect of the voltage probe is then to multiply every interference term in Eqs.~(\ref{eq:3onsev}-\ref{eq:3onsht}) with a factor $\sqrt{1-\varepsilon}$, where $\varepsilon$ is the probability to traverse the corresponding MZI arm \textit{without} entering the voltage probe. The maximal output power is then modified to
\begin{equation}
\label{eq:3maxpd}
		P_\text{max}=\frac{(\pi\kB\Delta T)^2}{8h}\frac{(1-\varepsilon)g^2(z)\sin^2\phi}{1+\sqrt{1-\varepsilon}s(z)\cos\phi}.
\end{equation}
Thus to a good approximation the power as well as the efficiency decrease linearly with increasing $\varepsilon$. Note that the power goes to zero for complete loss of phase-information, reflecting the fact that the delivered power is of quantum-mechanical origin only.

Since most experiments are not conducted at filling factor $\nu=1$ but at $\nu=2$, it is of experimental importance to investigate the effect of higher filling factors. At a QPC, only one of the edge channels can be partitioned at once. All others will be either fully transmitted or reflected. In order to fabricate a MZI, it must be the same edge channel which is partitioned at both QPCs. One can then see from Fig.~\ref{fig:setup1}, that a higher filling factor will only add connections with unit transmission between contacts $1$ and $3$ (and from contact $2$ back to itself). This is true independent of the choice for the partitioned channel. The charge current between the cold contacts, and with it the delivered power, is thus not affected but the heat current injected from contact $1$ is increased by a factor $\nu$. Therefore, the power remains unchanged while the efficiency drops down by a factor $\nu^{-1}$.

\subsection{\label{ssec:2terminal}Two-terminal MZI}

\begin{figure}
	\includegraphics[width=\columnwidth]{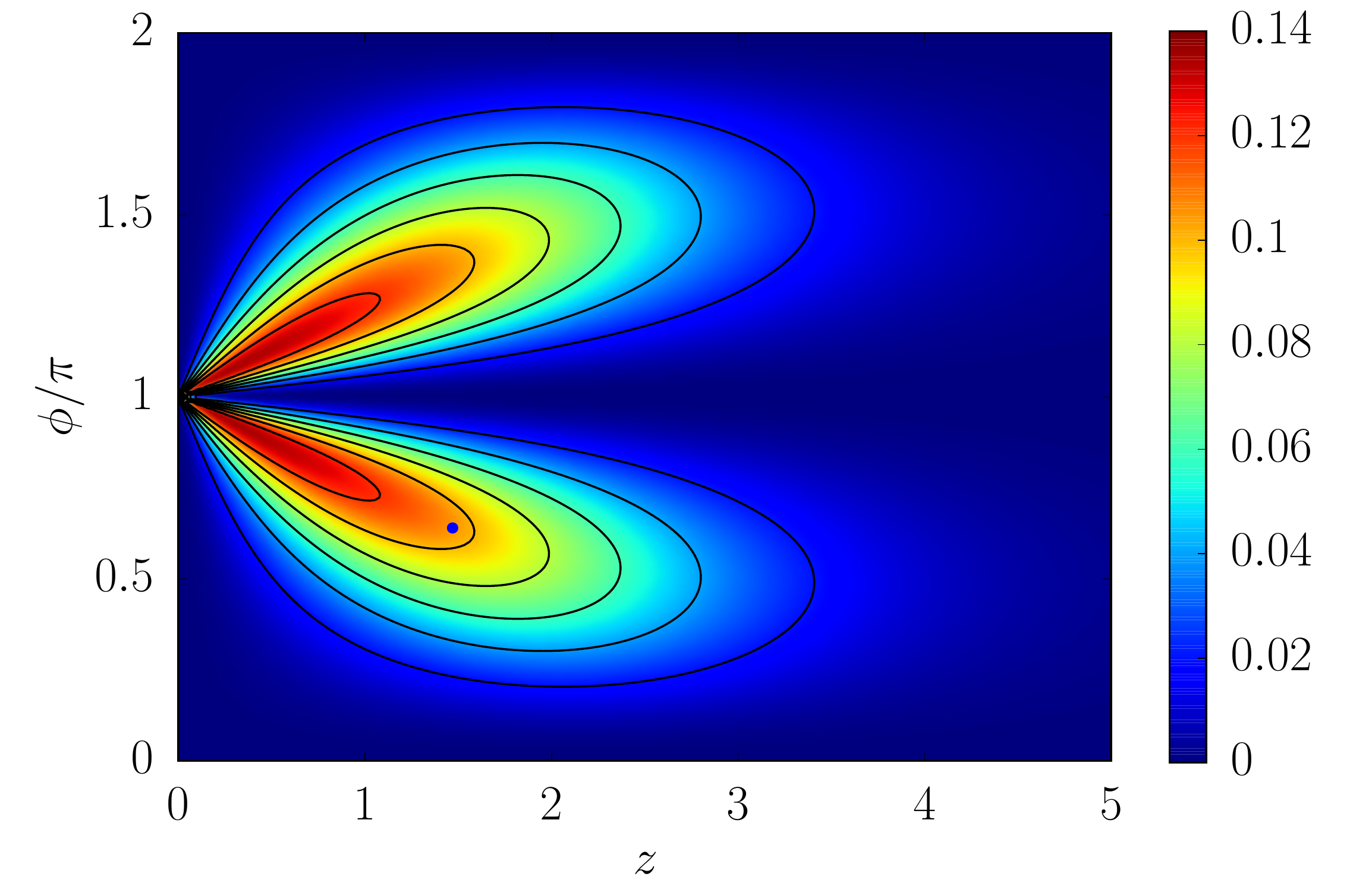}
	\caption{\label{fig:eff2terminal}Efficiency at maximum power for the effective two-terminal setup in units of $\eta_C$. Because of the interference contribution to the heat current, the efficiency is no longer directly proportional to the maximum power. The blue dot denotes the point where the power is maximized.}
\end{figure}

The three-terminal MZI can also be operated by demanding zero charge \textit{and} heat current in contact $1$, $I_1^e=I_1^h=0$, and applying a temperature bias between the remaining terminals. These boundary conditions are equivalent to the two-terminal setup one obtains by merging contacts $1$ and $3$ in Fig.~\ref{fig:setup1}. For that reason, we will refer to these boundary conditions as the two-terminal setup. The heat engine can be operated such that
\begin{equation}
\label{eq:2tvs}
\begin{aligned}
&T_1=T_3=T+\Delta T, \hspace{0.25cm}T_2=T,\hspace{0.25cm}V_1=V_3=0,\\
&I^e=I^e_3,\hspace{0.25cm}I^h=-I^h_3,\hspace{0.25cm}V_2=V.
\end{aligned}
\end{equation}
With this choice, the generated charge current remains the same as for the three-terminal setup. The Onsager coefficients $\mathcal{L}^{eV}$ and $\mathcal{L}^{eT}$ are thus given by Eqs.~(\ref{eq:3onsev},~\ref{eq:3onset}) and the Seebeck coefficient as well as the maximal power are unaffected by the choice of boundary conditions. The remaining two Onsager coefficients read
\begin{align}
	\label{eq:2onshv}
	\mathcal L^{hV}&=\frac{\pi (k_BT)^2}{2h}g(z)\sin\phi,\\
	\label{eq:2onsht}
	\mathcal L^{hT}&=\frac{\pi^2(\kBT)^3}{6h}\left(1+3[s(z)-2g(z)\coth(z)]\cos\phi\right).
\end{align}
Note that we now find $\mathcal{L}^{eT}=e\mathcal{L}^{hV}$, reflecting the fact that this setup does not effectively break the time-reversal symmetry (i.e., switching the magnetic field, the two remaining terminals and the path length difference $\tau$ results in the same setup). Furthermore, the heat current now obtains an interference contribution both from the temperature as well as from the voltage bias. We thus find that the thermoelectric properties of the MZI depend crucially on the choice of boundary conditions. The resulting efficiency at maximum power reads
\begin{equation}
\begin{aligned}
&\eta_{\rm maxP}=\eta_C\frac{3g^2(z)\sin^2\phi}{2+2s(z)\cos\phi}\times
\\&\bigg\{2+6\left[s(z)-2g(z)\coth(z)\right]\cos\phi -\frac{3g^2(z)\sin^2\phi}{1+s(z)\cos\phi}\bigg\}^{-1},
\end{aligned}
\end{equation}
and is plotted in Fig.~\ref{fig:eff2terminal}. 
At the point where the delivered power is maximized ($z\approx1.47$, $\phi\approx0.64\pi$), the efficiency reaches $\eta_{\rm maxP}\approx0.10\eta_C$ which is a clear improvement compared to the three-terminal setup. A slightly higher efficiency of $\eta_{\rm maxP}\approx0.14\eta_C$ can be reached for $z\rightarrow 0$ and $\phi\rightarrow\pi$. However, the power delivered by the heat engine vanishes for this configuration.

A further advantage of this setup is its insensitivity to higher filling factors. As can be inferred from Fig.~\ref{fig:setup1}, higher filling factors will only lead to edge states which return to the same contact they emerged from and therefore do not contribute to transport.

\subsection{\label{ssec:4terminal}Four-terminal double MZI}

\begin{figure}
	\includegraphics[width=\columnwidth]{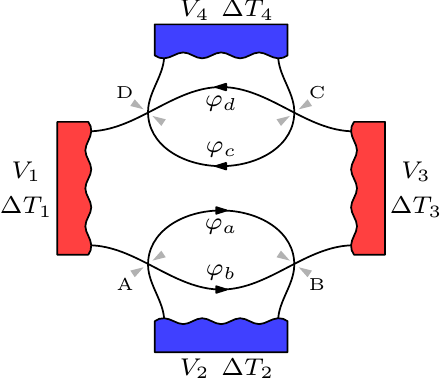}
	\caption{\label{fig:setup2}Schematic sketch of a four-terminal heat engine based on two electronic Mach-Zehnder interferometers. By heating contacts $1$ and $3$, a current is induced between contacts $2$ and $4$ due to the energy dependence of the transmission amplitudes. In this setup, every path traverses an MZI. Here $A$, $B$, $C$, and $D$ denote quantum-point contacts, $V_\alpha$ and $\Delta T_\alpha$ the bias voltage and the temperature bias applied at contact $\alpha$.}
\end{figure}

In this section, we discuss how the efficiency can be increased by using a setup based on two MZIs as sketched in Fig.~\ref{fig:setup2}. In this setup every path traverses a MZI leading to an interference contribution to all the charge and heat currents. This can be used to reduce the injected heat current and thus to increase the efficiency of the heat engine. Furthermore, as we will show, heating up two contacts instead of only one can lead to a gain in the output power by a factor bigger than two. Again restricting our investigation to half-transparent QPCs, where the interference effects are most pronounced, the non-zero scattering probabilities for this setup read 

\begin{equation}
\label{eq:4smat}
\begin{aligned}
&\left|S_{21}\right|^2=\left|S_{32}\right|^2=\frac{1}{2}\left(1+\cos\varphi_1\right),\\
&\left|S_{31}\right|^2=\left|S_{22}\right|^2=\frac{1}{2}\left(1-\cos\varphi_1\right),\\
&\left|S_{43}\right|^2=\left|S_{14}\right|^2=\frac{1}{2}\left(1+\cos\varphi_2\right),\\
&\left|S_{13}\right|^2=\left|S_{44}\right|^2=\frac{1}{2}\left(1-\cos\varphi_2\right),
\end{aligned}
\end{equation}
with
\begin{equation}
\label{eq:phasedoublemzi}
\varphi_i=\phi_i+\tau_iE/\hbar.
\end{equation}
Here, $\varphi_1=\varphi_a-\varphi_b$ denotes the lower, $\varphi_2=\varphi_c-\varphi_d$ the upper MZI. As for the three-terminal setup, we introduce a dimensionless quantity to quantify the path length difference
\begin{equation}
\label{eq:doublez}
z_i=\frac{\pi \kBT \tau_i}{\hbar}.
\end{equation}
We employ the four-terminal setup as a heat engine by heating contacts $1$ and $3$, demanding a vanishing charge current in the hot contacts. We thus have
\begin{equation}
\label{eq:4tvs}
\begin{aligned}
&T_1=T_3=T+\Delta T, \hspace{0.25cm}T_2=T_4=T,\\
&I^e=I^e_2,\hspace{0.25cm}I^h=-(I^h_1+I^h_3),\hspace{0.25cm}V_2=-V_4=-V/2.
\end{aligned}
\end{equation}
We remark that if we demand a fixed voltage instead of a vanishing charge current in the hot contacts, we obtain two copies of the two-terminal setup (i.e. double the power with the same efficiency).

\begin{figure}
	\includegraphics[width=\columnwidth]{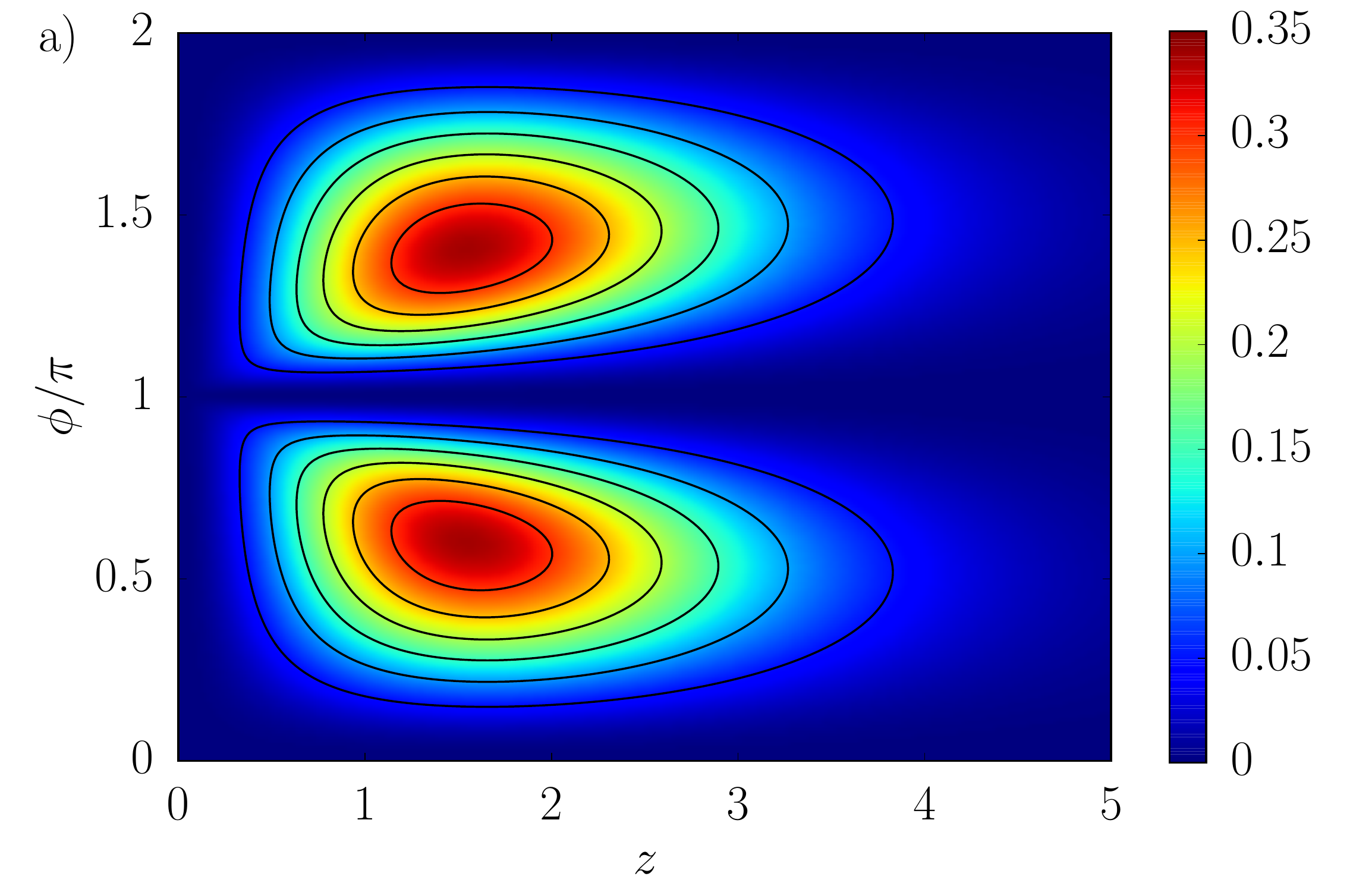}
	\includegraphics[width=\columnwidth]{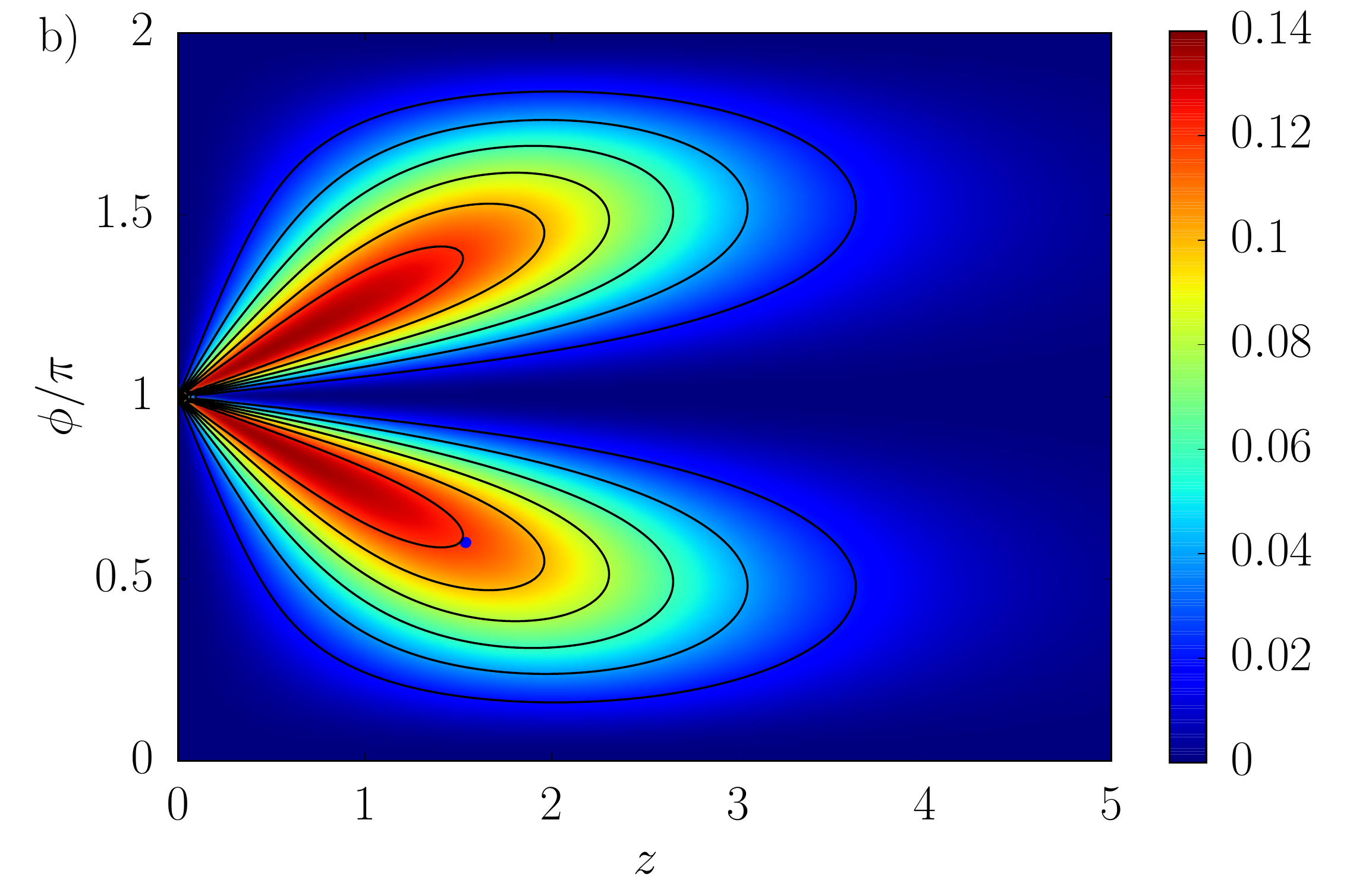}
	\caption{\label{fig:double_mzi_power_eta}Maximum power and efficiency at maximum power for the four-terminal double MZI setup at $\nu=1$. (a) Power in units of $(\kB\Delta T)^2/h$, (b) efficiency at maximum power in units of $\eta_C$. The blue dot denotes the point where the power is maximized.}
\end{figure}

Because most of the analytic expressions for the double MZI are rather cumbersome, we only show those expressions which reduce to an instructive form. For the off-diagonal Onsager coefficients we find $\mathcal{L}^{eT}=e\mathcal{L}^{hV}$, implying again that the four-terminal setup does not effectively break time reversal symmetry. The Seebeck coefficient is given by
\begin{equation}
\label{eq:doubleseebeck}
S=\frac{\pi k_B}{e}\left[\frac{g(z_2)\sin\phi_2}{1+s(z_2)\cos\phi_2}-\frac{g(z_1)\sin\phi_1}{1+s(z_1)\cos\phi_1}\right].
\end{equation}
For $z_1=-z_2$ and $\phi_1=\phi_2$, this is twice the value compared to the single MZI setups [cf.~Eq.~\eqref{eq:3seebeck}]. This is not surprising since in this setup twice as many terminals are heated. As it turns out, the condition $z=z_1=-z_2$ and $\phi=\phi_1=\phi_2$ also maximizes the delivered power. We therefore concentrate on this configuration which yields for the power
\begin{equation}
\label{eq:doublepower}
P_{\rm max}=\frac{(\pi k_B\Delta T)^2}{h}\frac{g^2(z)\sin^2\phi}{3+2s(z)\cos\phi-s^2(z)\cos^2\phi}.
\end{equation}
This expression reaches its maximum of $P_{\rm max}\approx 0.34(k_B\Delta T)^2/h=\unit[0.1]{pW/K^2}(\Delta T)^2$ at $z\approx1.54$ and $\phi\approx0.6\pi$. This is more than double the power obtained by a single MZI and comparable to heat engines based on resonant-tunneling quantum dots~\cite{jordan_powerful_2013}. The efficiency at maximum power reaches $\eta_{\rm maxP}\approx0.12\eta_C$ for the parameters that maximize the output power. This is a factor of three higher than the three-terminal setup. Similarly to the two-terminal case, we find a slightly higher efficiency, $\eta_{\rm maxP}\approx0.14\eta_C$ at $z_1=-z_2\rightarrow 0$ and $\phi_1=\phi_2=\pi$, a configuration where the delivered power vanishes. Both the maximum power as well as the efficiency at maximum power are plotted in Fig.~\ref{fig:double_mzi_power_eta}.

For the double MZI, a higher filling factor will introduce channels which connect the two hot contacts with unit scattering amplitude. %[cf.~Fig.~\ref{fig:setup2}]
Since we demand a vanishing charge current in the hot contact, these additional channels will move the chemical potentials of the hot contacts closer to each other. In contrast to the above setups, the double MZI is thus affected in a non-trivial way by a higher filling factor. For the case $\nu=2$, we find for the maximum power
\begin{equation}
\label{eq:doublepowernu2}
P_{\rm max}=\frac{(\pi k_B\Delta T)^2}{h}\frac{2g^2(z)\sin^2\phi}{7+6s(z)\cos\phi-s^2(z)\cos^2\phi},
\end{equation}
which reaches up to $P_{\rm max}\approx 0.30(k_B\Delta T)^2/h$ for $z\approx1.51$ and $\phi\approx0.62\pi$. For these parameters, the efficiency at maximum power reads $\eta_{\rm maxP}\approx0.11\eta_C$. We thus find that the power and the efficiency of the double MZI heat engine are only slightly reduced for $\nu=2$ which constitutes the filling factor at which most experiments are conducted.

\section{\label{sec:exp}Experimental feasibility}
A heat engine which is based on a quantum interference effect will naturally have a finite temperature window at which the temperatures are high enough to produce an appreciable current but low enough for the interference effect to survive. An additional constraint of our proposal is the energy dependence of the QPCs, which in principle decreases the interference effect. Note that this energy dependence can itself be exploited for thermoelectric effects~\cite{sanchez_chiral_2014}.
Describin the QPC by a saddle point potential~\cite{buttiker_quantized_1990} and assuming that the longitudinal and the transversal curvatures of the potential are equal and parametrized by the frequency $\omega$, the energy over which the QPC transmission changes is given by~\cite{buttiker_quantized_1990}
\begin{equation}
\label{eq:qpcen}
E_{QPC}=\frac{\hbar}{2\sqrt{2}}\left[\left(\omega_c^4+4\omega^4\right)^{1/2}-\omega_c^2\right]^{1/2},
\end{equation}
where $\omega_c=|eB/m^*|$ is the cyclotron frequency with the effective mass $m^*$. In order to be insensitive to the energy dependence of the QPC, the thermal energy $k_BT$ needs to be much smaller than $E_{QPC}$. In Ref.~\onlinecite{neder_unexpected_2006}, a magnetic field of $B=\unit[4.6]{T}$ was applied to reach a filling factor of $\nu=1$. Together with $m^*=0.067\,m_e$ and $\hbar\omega\approx \unit[2]{meV}$~\cite{van_wees_quantum_1991}, we find the condition $T\ll \unit[2]{K}$. Note that this condition relaxes to $T\ll \unit[2.9]{K}$ for $B=\unit[3.1]{T}$ which led to $\nu=2$ in Ref.~\onlinecite{neder_unexpected_2006}.

From the above analysis, we know that the proposed heat engine works best for $z=\pi k_BT\tau/\hbar\approx1.5$. Using a path length difference of $\Delta L\approx\unit[1.5]{\mu m}$~\cite{neder_unexpected_2006} and a drift velocity of $v_D\approx\unit[10^5]{m/2}$, we find the optimal temperature $T\approx \unit[240]{mK}$. The energy dependence of the QPCs should thus not spoil the performance of the proposed heat engine. However, the temperature bias that can be applied without leaving the linear response regime is limited. At the voltage that maximizes the power, a temperature bias of $\Delta T= \unit[60]{mK}$ leads to $I^e\approx\unit[0.044]{nA}$ for the three-terminal setup and $I^e\approx\unit[0.053]{nA}$ for the double MZI.

\section{\label{sec:Conclusions}Conclusions}
We investigated the thermoelectric properties of heat engines based on MZIs. The required energy dependence of the transmission probabilities arises from a length difference of the two interferometer arms and is thus due to an interference effect of purely quantum-mechanical nature. For the experimentally established three-terminal setup, we found a maximum power of $P_{\rm max}\approx\unit[0.04]{pW/K^2}(\Delta T)^2$ which is comparable to other proposals for mesoscopic heat engines \cite{sothmann_rectification_2012,jordan_powerful_2013}. With an efficiency at maximum power of $\eta_{\rm maxP}\approx0.042\eta_C$, this setup is however much below the theoretical bound of  $\eta_{\rm maxP}\leq\eta_C/4$~\cite{brandner_strong_2013}. The reason for this low efficiency is the scattering free connection from one of the drains to the source, making the injected heat current independent of any interference effect.

To decrease the injected heat, we investigated two modifications of the three-terminal setup. A two-terminal setup, which is equivalent to the three-terminal setup demanding zero charge and heat current in one of the contacts, and a four-terminal setup consisting of two MZIs. In both cases the efficiency increases because the injected heat current can be reduced by interference contributions. For the two-terminal setup, we found that the maximum power is insensitive to the change of boundary conditions but the efficiency increases to $\eta_{\rm maxP}\approx0.10\eta_C$. For the four-terminal setup, we find a maximum power of $P_{\rm max}\approx\unit[0.1]{pW/K^2}(\Delta T)^2$ with an efficiency of $\eta_{\rm maxP}\approx0.12\eta_C$. With this improvement, our proposed device comprises a powerful and efficient heat engine.

Additionally, we investigated the effect of dephasing, higher filling factors and the temperature range at which we expect our heat engine to perform optimally. We find that the proposed device performs well under realistic conditions, confirming that finite temperatures can give rise to thermoelectric effects of purely quantum-mechanical nature.\\

\acknowledgments

This work was supported by the Swiss NSF and the COST Action MP1209. We thank C. Flindt and R. S\'anchez for stimulating discussions. PH gratefully acknowledges the hospitality of the McGill University, where part of the work was done.

% \bibliographystyle{apsrev4-1}
% \bibliography{Meine_Bibliothek}


\end{document}